\documentclass[10pt]{iopart}
\newcommand{\eqref}[1]{\eref{#1}}
\usepackage[T1]{fontenc}
\usepackage{ae} 
\usepackage{amssymb}
\usepackage{amsfonts} 
\usepackage{mathrsfs}
\usepackage[disable]{todonotes}%

\usepackage{units}
\newcommand{\tallrow}{\\[-0.8em]}
\newcommand{\Eceff}{$E_{\rm c}^{\rm eff}$}
\newcommand{\Eceffeq}{E_{\rm c}^{\rm eff}}
\newcommand{\Ectot}{E_{\rm c}^{\rm tot}}
\newcommand{\taur}{\tau_{\rm syn}}
\newcommand{\taurInv}{\taur^{-1}}
\newcommand{\nofrac}[2]{#1/#2}
\newcommand{\aj}{\bar{a}_j}
\newcommand{\IjInv}{I_j^{\scalebox{0.5}[0.7]{\( - \)}\!1}}
\newcommand{\nuSD}{\bar{\nu}_{\rm s0}}
\newcommand{\nuSH}{\bar{\nu}_{\rm s1}}
\newcommand{\nuDB}{\bar{\nu}_{\rm D0}}
\newcommand{\nuDC}{\bar{\nu}_{\rm D1}}
\newcommand{\lnLStar}{\ln\Lambda_{\rm c}}
\newcommand{\CFP}{C_{\rm FP}\{f\}}
\newcommand{\Fbr}{F_{\rm br}}
\newcommand{\x}{x_{\rm rad}}
\newcommand{\pcStar}{p_{\rm c}^\star}
\newcommand{\pcStarO}{p_{\rm c 0}^\star}
\newcommand{\phibrI}{\phi_{\rm br 0}}
\newcommand{\phibrII}{\phi_{\rm br 1}}

\usepackage[T1]{fontenc}
\usepackage{bm}
\usepackage{color}
\usepackage{graphicx}
\usepackage[breaklinks=true]{hyperref}
\hypersetup{
  unicode=false,          % non-Latin characters in Acrobat’s bookmarks
  pdftoolbar=true,        % show Acrobat’s toolbar?
  pdfmenubar=true,        % show Acrobat’s menu?
  pdffitwindow=false,     % window fit to page when opened
  pdfstartview={FitH},    % fits the width of the page to the window
  pdftitle={Effect of partially ionized impurities on the effective critical electric field},    % title
  pdfauthor={L Hesslow, O Embréus, G J Wilkie, G Papp and T Fülöp},     % author
  pdfnewwindow=true,      % links in new PDF window
  colorlinks=true,        % false: boxed links; true: colored links
  linkcolor=blue,         % color of internal links (change box color with linkbordercolor)
  citecolor=blue,         % color of links to bibliography
  filecolor=blue,         % color of file links
  urlcolor=blue           % color of external links
}
\begin{document}
\title[]{Effect of partially ionized impurities and radiation on the effective critical electric field for runaway generation}

\author{L~Hesslow$^1$, O~Embr\'eus$^1$, G J Wilkie$^1$, G~Papp$^2$ and T~F\"ul\"op$^1$}
\address{$^1$Department of Physics, Chalmers University of Technology, 41296 Gothenburg, Sweden}
\address{$^2$Max-Planck-Institute for Plasma Physics, D-85748 Garching, Germany}
\ead{hesslow@chalmers.se}

\vspace{10pt}
\begin{indented}
\item[] (Dated: \today)
\end{indented}

\begin{abstract}
  We derive a formula for the effective critical electric field for runaway
  generation and decay that accounts for the presence of partially
  ionized impurities in combination with synchrotron and
  bremsstrahlung radiation losses.  We show that the effective critical field is drastically larger than the classical Connor--Hastie field, and even exceeds the value obtained by replacing the free electron density by the total electron density (including both free and bound electrons).
Using a
  kinetic equation solver with an inductive electric field, we show
  that the runaway current decay after an impurity injection is expected
  to be linear in time  and proportional to the effective critical electric
  field in highly inductive tokamak devices. This is relevant for the
  efficacy of mitigation strategies for runaway electrons since it reduces the required amount of injected impurities to achieve a certain current decay rate.
 \end{abstract}
\noindent{\it Keywords\/}: runaway electron, tokamak,  disruption, Fokker--Planck

\submitto{\PPCF}
%\maketitle
%\ioptwocol

%%%%%%%%%%%%%%%%%%%%%%%%%%%%%%%%%%%%%%%%%%%%%%%%%%%%%%%%%%%

\section{Introduction}

When a plasma carrying a large electric current is suddenly cooled, as happens in tokamak disruptions, a large toroidal electric field is induced due to
the dramatic increase of the plasma resistivity. If this electric
field is larger than a certain critical electric field, a relativistic
runaway electron beam can be generated~\cite{Dreicer1960,connor}. Such
runaway beams can damage the plasma facing components on impact due to
localized energy deposition. Therefore, runaway electrons constitute a
significant threat to large tokamak experiments (e.g.~ITER)~\cite{Reux2015,HollmannDMS,Boozer2015}.

To minimize the risk of damage, it is crucial to understand the
runaway-electron dynamics.
{Disruption mitigation by material injection is motivated by the strong influence of partially ionized atoms, as observed in experiments~\cite{Reux2015,HollmannDMS,pautasso17disruption}. 
It is therefore important to have accurate
models of the interaction between fast electrons and the partially
screened nuclei of heavy ions.  Fast electrons are not simply
deflected by the Coulomb interaction with the net charge of the ion,
but probe its internal electron structure, so that the nuclear
charge is not completely screened. Energetic electrons can therefore be expected to
experience higher collision rates against partially ionized impurities compared to a fully ionized plasma with the same effective charge, leading to a
more efficient damping. There has been a considerable effort to produce a detailed theoretical description of this process~\cite{Kirillov,zhogolev,Hesslow,BA2017}.

A recent paper presented a generalized collision operator which describes the interaction between fast electrons and partially screened impurities via analytic modifications to the collision frequencies~\cite{Hesslow}.
 The elastic electron-ion
collisions were modeled quantum-mechanically in the Born
approximation as in~\cite{Kirillov,zhogolev}, however, to obtain the required electron-density distribution of the impurity ions~\cite{Kirillov,zhogolev} used the Thomas-Fermi model. In Ref.~\cite{Hesslow} we used fitted results from density functional theory (DFT) thereby providing a more accurate description.  To
describe inelastic collisions with bound electrons, we employed
Bethe's theory for the collisional stopping power~\cite{bethe}, with
mean ionization energies for ions calculated in~\cite{sauer2015}.
Our results show that, already at sub-relativistic electron energies, the deflection and slowing-down frequencies are
increased significantly compared to standard collisional theory~\cite{Hesslow}.

The quantity that is arguably the most important for runaway
generation and decay is the threshold, or critical, electric field, which in a fully
ionized plasma without radiation losses is given by the Connor-Hastie
field $E_{\rm c}\!=\!n_{\rm e} e^3 \ln\Lambda/(4 \pi \epsilon_0^3 m_{\rm e}
c^2)$~\cite{connor}, where $n_{\rm e}$ and $m_{\rm e}$ are the electron density and
mass, $\ln\Lambda$ is the Coulomb logarithm, $\epsilon_0$ is the
vacuum permittivity and $c$ is the speed of light. Below the
threshold field no new runaway electrons are produced and all
preexisting runaways eventually thermalize.  There is a wealth
of experimental evidence that the critical electric field is much
higher than $E_{\rm c}$ given above~\cite{Granetz,Hollmann2013,MartinSolis,Paz-Soldan,Popovic,Plyusnin2018}. Well-diagnosed
and reproducible experiments in quiescent plasmas on a wide range of
tokamaks show that measured threshold electric fields can be approximately an order of magnitude 
higher than predicted by the Connor-Hastie threshold~\cite{Granetz,Plyusnin2018}. 
Furthermore, it has been shown that the runaway
electron current decays much faster after high-$Z$ particle injection than expected from conventional
theory~\cite{connor}, in contrast to low-$Z$
particle injection which results in a current decay rate only slightly
below that expected~\cite{Hollmann2013}. From a theoretical point
of view, the threshold electric field is expected to be higher than
$E_{\rm c}$, as can be influenced by
synchrotron~\cite{Stahl2015,AleynikovPRL} and bremsstrahlung radiation
losses, and also, as we will show here, by the presence of partially
ionized atoms. The value of the critical electric field is not only interesting theoretically -- it is of immense practical importance as
it determines the amount of material that has to be injected in
disruption mitigation schemes~\cite{breizman2014}.

In this paper we derive an analytical expression for the effective
critical field for runaway generation and decay that takes into
account the presence of partially screened impurities, using the
generalized collision operator derived in~\cite{Hesslow}. We present a formula that accounts for
arbitrary ion species in combination with synchrotron and
bremsstrahlung losses.  We show that the effect of partially screened
impurities is captured by replacing the plasma density in the
critical electric field with an effective density $n\! =\! n_{\rm free}\! + \!\kappa n_{\rm bound}$, where $\kappa$ is typically in the range 1-2 which implies that the effect of bound electrons is significantly larger than suggested by previous studies~\cite{rosenbluthPutvinski}.  Furthermore, using a kinetic equation solver
with a 0D inductive electric field, we verify the prediction from~\cite{breizman2014}, that the runaway current
in highly inductive tokamak devices after impurity injection will decay linearly with time at a rate proportional to the effective
electric field. 
We expect these findings will facilitate future comparisons with experimental observations of runaway-current decay, however such analysis is beyond the scope of the present paper.

The structure of the paper is as follows. In section~\ref{sec:coll} we
describe the kinetic model accounting for the effect of partial
screening in both the generalized collision operator and the
bremsstrahlung operator.  Then we proceed in section~\ref{sec:ecrit} to
derive analytical expressions for the effective critical electric
field in the presence of partially ionized impurities. This
calculation generalizes the results in~\cite{AleynikovPRL}, in which the critical electric field 
was calculated by assuming rapid pitch-angle dynamics in the
Fokker--Planck equation. In contrast to~\cite{AleynikovPRL}, 
our study includes the effect of
partially ionized impurities and bremsstrahlung losses. We demonstrate
how the presence of partially screened impurities affects both
synchrotron losses (through pitch-angle scattering) and bremsstrahlung
(as partial screening affects the bremsstrahlung cross-section). In
section~\ref{sec:decay} we discuss the decay of a  runaway current when heavy
impurities are injected. Through kinetic simulations, we demonstrate the accuracy of the
analytical expressions for the effective critical electric field and the
current decay.  Finally in section~\ref{sec:concl} we summarize our
conclusions.

\section{Kinetic equation including partially screened impurities}
\label{sec:coll}
In a uniform, magnetized plasma, the kinetic equation for relativistic electrons can be written as follows:
\begin{eqnarray}
\fl \frac{\partial f}{\partial \tau} + \underbrace{\frac{E}{E_{\rm c}}\!\left(\xi \frac{\partial f }{\partial p} + \frac{1\!-\!\xi^2}{p}\frac{\partial f}{\partial \xi}\right)}_{\rm electric\, field} \nonumber 
\\ =  \underbrace{\vphantom{\frac{\partial}{\partial p}}\CFP + S_{\rm ava}}_{\rm collisions}+ \underbrace {  C_{\rm br}\{f\}-\frac{\partial}{\partial \bi{p}}\!\cdot\! \left(\bi{F}_{\rm syn} f\right)}_{\rm radiation \, reaction}
 \label{eq:FP}\,,
\end{eqnarray}  
where $f$ is the electron distribution function, $\CFP$ is the
 partially screened Fokker--Planck collision operator as described in
 section~\ref{subsec:collfreq}, which accounts for ionizing as well as elastic collisions.
 The avalanche source is denoted $S_{\rm ava}$ and $E$ is the component of the electric field which is
 antiparallel to the magnetic field $\bi{B}$.  Radiation losses
are modeled by $C_{\rm br}$ (the bremsstrahlung collision operator)
 and $\bi{F}_{\rm syn}$ (the synchrotron radiation reaction
 force), which are described in section~\ref{subsec:rad}.  The normalized
 momentum is defined as $p\!=\!\gamma v/c$ is (with $\gamma$ the
 Lorentz factor), $\xi = \bi{p}\cdot\bi{B}/(pB)$ is the cosine of the pitch-angle, and the time
 variable $\tau$ is normalized to the relativistic collision time
 \begin{equation*}\tau_{\rm c} \!=\! \nofrac{4 \pi \epsilon_0^2 m_{\rm e}^2 c^3}{(n_{\rm e} e^4
   \lnLStar)},
   \end{equation*} 
   where we introduced a relativistic Coulomb logarithm
\begin{equation}
 \lnLStar =  \ln\Lambda_0  +  \frac{1}{2}\ln\frac{m_{\rm e} c^2}{T}\approx 14.6+0.5 \ln (T_{\rm eV}/n_{e20}).
 \label{eq:lnLStar}
\end{equation}
Here, $T_{\rm eV}$ is the temperature in $\rm eV$, $n_{e20}$ is normalized to $\unit[10^{20}]{m^{-3}}$ and $\ln\Lambda_0 = 14.9-0.5 \ln n_{e20}+\ln T_{\rm keV}$ is the thermal electron-electron Coulomb logarithm~\cite{wesson}.
The temperature dependence of $\lnLStar$ is reduced compared to $\ln\Lambda_0$ as it describes collisions between thermal particles and relativistic electrons; 
\eqref{eq:lnLStar}~corresponds to evaluating the energy-dependent electron-ion Coulomb logarithm $\ln\Lambda^{\rm ee}$ at $\gamma = 2$. 
%The superthermal Coulomb logarithms, which multiply the collision frequencies for electrons with ions and free electrons, are determined by the logarithm of the Debye length divided by the respective de Broglie wavelengths. 
For future reference, the superthermal Coulomb logarithms are given by~\cite{SolodovBetti} 
\begin{equation}
\ln \Lambda^{\rm ee} = \ln\Lambda_{\rm c} + \ln\sqrt{(\gamma\!-\!1)} \label{eq:lnLee}
\end{equation} and 
\begin{equation}
\ln \Lambda^{\rm ei} = \ln\Lambda_{\rm c} + \ln (\sqrt{2} p)\,.
\end{equation}

The parallel electric field $E$ is thus most naturally compared to the  critical electric field $E_{\rm c}$ defined with the relativistic Coulomb logarithm $\lnLStar$ (rather than the thermal $\ln\Lambda_0$):
 \begin{equation*} E_{\rm c} = \frac{n_{\rm e} e^3 \lnLStar}{4 \pi \epsilon_0^2 m_{\rm e} c^2}=\frac{m_{\rm e} c}{e\tau_{\rm c}} .\end{equation*} 

\subsection{Collision frequencies with partially ionized impurities}
\label{subsec:collfreq}
When acting on relativistic electrons and $T\ll m_{\rm e} c^2$, the linearized Fokker--Planck collision operator $\CFP$ can be simplified to
\begin{equation*}
\CFP = \nu_{\rm D} \mathscr{L} \{f\} +
\frac{1}{p^2}\frac{\partial}{\partial p}\left(p^3\!  {
      \nu_{\rm s}} f \right),
\end{equation*}
where $\mathscr{L} = \frac{1}{2}\frac{\partial}{\partial \xi}\left( 1-\xi^2\right)\frac{\partial}{\partial \xi}$ is the Lorentz scattering operator. 
The 
slowing-down frequency $\nu_{\rm s} = \nu_{\rm s}^{\rm ee}$ and the deflection frequency $\nu_{\rm D} = \nu_{\rm D}^{\rm ee}+\nu_{\rm D}^{\rm ei}$ are well known in the limits of
{\em complete screening} (i.e.~the electron interacts only with the
net ion charge) and {\em no screening} (the electron experiences the
full nuclear charge). The generalized expressions for $\nu_{\rm D}^{\rm ei}$ and
$\nu_{\rm s}^{\rm ee}$ taking into account partial screening are given in~\cite{Hesslow}.

Focusing on the effective critical electric field $E_{\rm c}^{\rm eff}$ in this
paper, the following equations are specialized to the superthermal
momentum region, in which the critical momentum $p_{\rm c}$ corresponding to
$E_{\rm c}^{\rm eff}$ is found. Thus all of the following expressions are
given for superthermal electrons.  

The generalized deflection frequency is, in units of $\tau_{\rm c}^{-1}$, given by

\begin{eqnarray} 
\fl \nu_{\rm D} =\frac{\gamma}{p^3}\bar{\nu}_{\rm D},\nonumber\\
\fl \bar{\nu}_{\rm D} = \frac{1}{ \lnLStar}\bigg[
\ln\Lambda^{\rm ee} + \ln\Lambda^{\rm ei} Z_{\rm eff} \nonumber\\
+\sum_{j}\frac{n_j}{n_{\rm e}} \left((Z_j^2\!-\!Z_{0,j}^2)\ln\left(
    \aj p\right)-\frac{2}{3}N_{{\rm e},j}^2\right)
\bigg].\label{eq:nuD}
\end{eqnarray}
Here, $Z_{0,j}$ is the ionization state, $Z_j$ is the charge number and $N_{{\rm e},j} = Z_j\!-\!Z_{0,j}$ is the number of bound electrons of the nucleus for species $j$, 
 $Z_\mathrm{eff} = \sum_j n_j Z_{0,j}^2/n_{\rm e}$,
where $n_j$ is the density of species $j$, and $n_{\rm e}$ represents the
density of free electrons. The parameter $\aj$ was determined from DFT
calculations, and is an effective ion size which depends on the ion
species $j$. These constants are given for argon and neon in
table~\ref{tab:constants} in~\ref{app:const}. In~\eqref{eq:nuD}, we have assumed $p \gg 1/\bar{a}_j \simeq 10^{-2}$. Figure~\ref{fig:Nu}a shows the enhancement of the deflection frequency
for singly ionized argon and neon. At typical runaway energies in the MeV range, the
enhancement is more than an order of magnitude compared to taking the
limit of \emph{complete screening} and neglecting the variation of the
Coulomb logarithm, which would give $\bar{\nu}_{\rm D} = 1+Z_{\rm eff}$.

In the limit of $p\gg1$, the deflection frequency~\eqref{eq:nuD} can be approximated
by 
\begin{equation}
\bar{\nu}_{\rm D}\approx (\nuDB+\nuDC\ln p) \label{eq:nuDApprox}
\end{equation}
 where the constants are given by
 %\numparts
\begin{eqnarray}
\nuDB =  1\!+\!Z_{\rm eff} \!+\!
   \frac{1}{\lnLStar\!}\!\sum_{j}\!\frac{n_j}{n_{\rm e}}\!\left(\!
  (Z_j^2\!-\!Z_{0,j}^2)\!\ln\aj\!-\!\frac{2}{3}N_{{\rm e},j}^2\right)\!, \,\quad \label{eq:nuDB}
  \\
  \nuDC = \frac{1}{\lnLStar} \sum_{j}\frac{n_j}{n_{\rm e}} Z_j^2  \label{eq:nuDC}.
  \end{eqnarray}
 % \endnumparts

 For the superthermal \begin{figure}
 \centering
  \includegraphics[width=(0.3\linewidth+0.3\textwidth)]{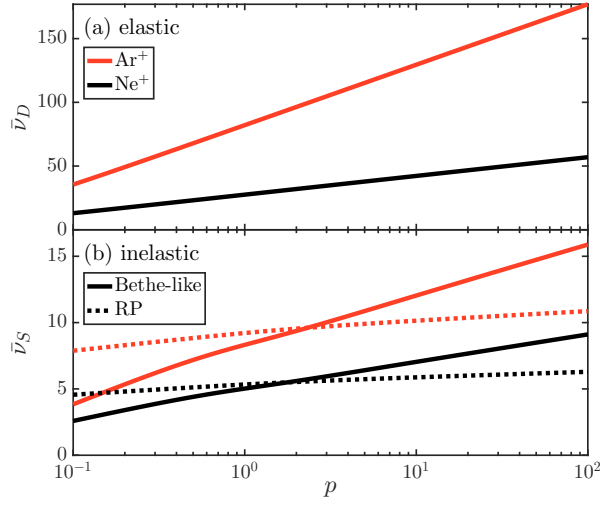}
 \caption{\label{fig:Nu} \small \noindent (a) The deflection frequency
   and (b) the slowing-down frequency as a function of the
   incoming-electron momentum, for both ${\rm Ar}^+,$ (black) and  ${\rm      Ne}^+,$ (red). These are normalized such that $\nu_{\rm D} =\tau_{\rm c}^{-1}(\nofrac{\gamma}{p^3})\bar{\nu}_{\rm D}$ and $\nu_{\rm s} =\tau_{\rm c}^{-1}(\nofrac{\gamma^2}{p^3}) \bar{\nu}_{\rm s}$.  
   The solid lines denote $\nu_{\rm D}$ from \eqref{eq:nuD} and $\nu_{\rm s}$ from \eqref{eq:nuS}, respectively. The
   approximate Rosenbluth-Putvinski (RP) model of $\nu_{\rm s}$~\cite{rosenbluthPutvinski} is
   shown in dotted line.  Parameters: $T=\unit[10]{eV}$ and  $n_Z=n_{\rm e} = \unit[10^{20}]{m^{-3}}$. }
\end{figure}
slowing-down frequency, we obtain, in units of $\tau_{\rm c}^{-1}$,
\begin{eqnarray}
 \nu_{\rm s} &=\frac{\gamma^2}{p^3} \bar{\nu}_{\rm s}\,, \nonumber\\
\bar{\nu}_{\rm s} &= \frac{1}{\lnLStar}\bigg(\! \ln\Lambda^{\rm ee}\!+
\!\sum_j\! \frac{n_j}{n_{\rm e}}N_{{\rm e},j} \left( \ln h_j -\beta^2\right) \!\!\bigg).
\label{eq:nuS}
\end{eqnarray}
Here, $h_j = p\sqrt{\gamma-1}/I_j$ and $I_j$ is the mean excitation
energy of the ion, normalized to the electron rest energy~\cite{sauer2015}; see table~\ref{tab:constants} in~\ref{app:const}. As $\nu_{\rm s}$ given in~\eqref{eq:nuS} is
based on the Bethe stopping-power formula matched to the low-energy
asymptote corresponding to complete screening, we refer to it as the
\emph{Bethe-like} model.  As shown in figure~\ref{fig:Nu}b, the slowing-down
frequency is enhanced significantly compared to the completely
screened limit with constant Coulomb logarithm, where $\bar{\nu}_{\rm s} =
1$.  The enhancement is also significantly different from a widely
used rule of thumb that is mentioned in passing by Rosenbluth and
Putvinski~\cite{rosenbluthPutvinski}, which suggests that inelastic
collisions with bound electrons can be taken into account by adding
half the number of bound electrons to the number of free electrons.
As shown in figure~\ref{fig:Nu}, the Rosenbluth-Putvinski (RP) model
overestimates the slowing-down frequency at low energies and is a
significant underestimation at high runaway energies. The weak
energy-dependence of the RP model is due to the energy-dependence in
the electron-electron Coulomb logarithm in~\eqref{eq:lnLee}.

In the ultra-relativistic limit $p\gg 1$, the slowing-down frequency
~\eqref{eq:nuS} is approximately
\begin{equation} 
\bar{\nu}_{\rm s} \approx (\nuSD + \nuSH \ln p)\label{eq:nuSApprox},
\end{equation}
where
%\numparts
  \begin{eqnarray}
\nuSD&= 1+\frac{1}{\lnLStar}\!\sum_j\!
\frac{n_j}{n_{\rm e} }N_{{\rm e},j}\left(\ln \IjInv \!-\!1\right) \label{eq:nuSD},\\
\nuSH &= \frac{1}{2}\frac{1}{\lnLStar}\bigg(1+\!\sum_j 3\,\frac{ n_j }{n_{\rm e}
}N_{{\rm e},j}\bigg).\label{eq:nuSH}
\end{eqnarray}
%\endnumparts

\subsection{Radiation losses}
\label{subsec:rad}
At the high densities typical of post-disruption scenarios,
bremsstrahlung may be an important energy loss mechanism compared to
synchrotron radiation
reaction~\cite{BakhtiariPRL2005,EmbreusBrems2016}. In a fully ionized
plasma, the required density for bremsstrahlung dominance
is~\cite{EmbreusBremsOld}
\begin{equation}
n_{\rm e,20} \gtrsim B_{\rm T}^2\,,
\label{eq:synchbrems}
\end{equation}
with $B_{\rm T}$ in units of Tesla and $n_{\rm e, 20}$ normalized to
 $\unit[10^{20}]{m^{-3}}$. 
 In a partially ionized plasma, both bremsstrahlung and synchrotron
 losses will be enhanced, the latter through the increased pitch-angle
 scattering. Both radiative energy loss channels can therefore be
 significant at densities characteristic of disruptions and are
 included in this paper.

The synchrotron radiation reaction force is given by~\cite{hirvijokiALD2015,HirvijokiBump2015}
\begin{eqnarray}
\fl \frac{\partial}{\partial \bi{p}}\!\cdot\! \left(\bi{F}_{\rm syn} f\right) =
 -\!\frac{1}{p^2}\frac{\partial}{\partial p} \!
\left( \frac{p^3\gamma}{\taur} (1\!-\!\xi^2){f} \right) 
\nonumber\\ 
+ 
\frac{\partial}{\partial \xi}\left(\frac{\xi(1\!-\!\xi^2)}{\taur \gamma}{f}\right),
\label{eq:Frad}
\end{eqnarray}
where $\taur$ is the synchrotron radiation-damping timescale normalized to $\tau_{\rm c}$:
 \begin{equation}
 \taurInv =  \frac{\tau_{\rm c} e^4 B^2}{6 \pi \epsilon_0 m_{\rm e}^3 c^3}
 \approx \frac{1}{15.44\lnLStar}\frac{ B_{\rm T}^2 }{  n_{\rm e, 20}  }.
 \label{eq:taurInv}
 \end{equation}

We model partially screened bremsstrahlung with a Boltzmann operator
as presented in~\cite{EmbreusBrems2016}, using the model
that neglects the angular deflection due to the bremsstrahlung
process:
\begin{eqnarray*}
\fl C_{\rm br}(p,\xi) =  \int \! v_1 f(p_1,\xi)\frac{\partial {\sigma}^{\rm br}(p,p_1)}{\partial p}\mathrm{d}p_1 
 \\
- v f(p,\xi)\sigma^{\rm br}(p),
\end{eqnarray*} 
where $\partial \sigma^{\rm br}(p,p_1)/\partial p$ is the normalized cross-section for an incident electron with momentum $p_1$ to end up with momentum $p$ after emitting a bremsstrahlung photon carrying the energy difference, and $\sigma^{\rm br}$ is the total bremsstrahlung cross section for an incident electron of momentum $p$. The integration is taken over $\sqrt{(\gamma + k_\mathrm{c})^2-1} \leq p_1 <\infty$, where, following~\cite{EmbreusBrems2016}, photon energies are cut off at $0.1\%$ of the kinetic energy of the outgoing electrons in order to resolve the infrared divergence, i.e.~$k_c = (\gamma-1)/1000$. The partially screened bremsstrahlung cross section is
given in~\cite{Koch1959, Seltzer1985}:
\begin{eqnarray}
\fl \frac{\partial{\sigma}^{\rm br}}{\partial p}(p,p_1) = \frac{\alpha}{\pi \ln\Lambda_{\rm c}} \frac{1}{k} \sum_j \frac{n_j}{n_{\rm e}}\left[
\left(1+\frac{\gamma^2}{\gamma_1^2}\right)
\right. \nonumber \\ \times
\left( Z_j^2 + 
\int_{q_0}^1 [Z_j\!-\!F_j(q)]^2 \frac{(q-q_0)^2}{q^3} \mathrm{d}q\right)
\nonumber \\ 
-\frac{2}{3}\frac{\gamma}{\gamma_1}\left(\frac{5}{6}Z_j^2
+\int_{q_0}^1 [Z_j\!-\!F_j(q)]^2
\right. \nonumber \\  \left.\left. \times
 \frac{q^3+3q q_0^2[1-2q q_0^2\ln(q/q_0)]-4q_0^3}{q^4}  \mathrm{d}q\right) \right]
\label{eq:sigmaBrems}
\end{eqnarray}
where $k$ is the photon momentum and $q_0 = p_1-p-k$.  We use the form factor $F(q)$ for partially ionized
atoms presented in~\cite{Hesslow},
\begin{equation*}
F_{j}(q) = \frac{N_{{\rm e},j}}{1+(q \aj)^{3/2}}\,.
\end{equation*}
 
 In order to get an analytically tractable problem when deriving the
 effective critical electric field, a simplified bremsstrahlung mean-force
 stopping power will be used in section~\ref{sec:ecrit}.  Although a
 mean-force model has been shown to significantly alter the
 steady-state electron distribution compared to the full Boltzmann
 model, it captures the mean energy
 accurately~\cite{EmbreusBrems2016}, and is therefore sufficient for
 the purpose of deriving the effective critical electric field. This assumption
 is verified with numerical calculations using the full Boltzmann
 operator in section~\ref{sec:decay}.
 
 For the mean force model, we have
 \begin{equation}
C_{\rm br}\{f\} \approx - \frac{\partial}{\partial \bi{p}}\!\cdot\! \left(\bi{F}_{\rm br} f\right) =  \!\frac{1}{p^2}\frac{\partial}{\partial p} \!
\left( p^2 \Fbr f \right),
\label{eq:Cbr}
 \end{equation}
 where the bremsstrahlung mean force is given by $F_{\rm br}(p) = \int \, k (\partial \sigma^{\rm br}(p_1,p)/\partial p_1) \,\mathrm{d} p_1$, the integral taken over all allowed outgoing momenta $p_1$.
 For argon and neon, a numerical investigation of~\eqref{eq:sigmaBrems} shows that $\Fbr$ is well approximated by
 \begin{eqnarray}
 \Fbr &\approx p(\phibrI + \phibrII\ln p) 
 \nonumber \\
 &\equiv \frac{p \alpha}{\lnLStar}\sum_j \frac{n_j}{n_{\rm e}} Z_j^2(0.35 + 0.20 \ln p)  
 \label{eq:phiBrems}.
 \end{eqnarray} 
\section{Effective critical electric field}
\label{sec:ecrit}
\indent The critical electric field is a central parameter for both
generation of a runaway current and for its decay rate in a highly
inductive tokamak; in the latter case, it is predicted that once the
Ohmic current has dissipated, the induced electric field will be close
to the critical electric field so that the current decays according to
${\rm d}I/{\rm d} t\!=\!2 \pi R E_{c}^{\rm eff}/L$~\cite{breizman2014}, where
$L\!\sim\!\mu_0 R$ is the self-inductance and $R$ is the major radius
of the tokamak. The physical argument is that the runaway avalanche
timescale is much faster than the inductive timescale, and therefore
the electric field must be close to the critical electric field to
prevent rapid current variations.

We calculate the effective electrical field due to collisions with
partially screened ions by finding the minimum electric field
\Eceff\ that satisfies the pitch-angle averaged force-balance equation
\begin{equation*}
\langle e E \xi - F    \rangle = 0\,,
\end{equation*}
where $F$ denotes the collisional and radiation forces on a runaway electron. 

In order to find \Eceff, we assume rapid pitch-angle dynamics compared to the timescale 
of the energy dynamics~\cite{lehtinen,AleynikovPRL}. In the
kinetic equation~\eqref{eq:FP}, this amounts to requiring that the
pitch-angle flux vanishes. Since $\taurInv\ll 1$ from \eqref{eq:taurInv}, we can neglect the effect of radiation on the pitch-angle distribution (term marked as ``neglect'' below) as well as the effect of the avalanche source, which is slower than both pitch-angle scattering and collisional friction. We demonstrate the validity of these assumptions in~\ref{app:dist} by comparing the resulting critical electric field and angular distribution to kinetic simulations. Inserting the collision frequencies~\eqref{eq:nuDApprox} and \eqref{eq:nuSApprox} as well as the radiation terms~\eqref{eq:Frad} and \eqref{eq:Cbr},
the kinetic equation~\eqref{eq:FP} can be rewritten
\begin{eqnarray}
\fl \frac{\partial \bar{f}}{\partial \tau} = \frac{\partial}{\partial p} 
\left[\left(- \frac{\xi E}{E_{\rm c}} + p\nu_{\rm s}+ \Fbr+ \frac{p\gamma}{\taur} (1\!-\!\xi^2)\right)\bar{f} \right] \nonumber \\ 
+\frac{\partial}{\partial \xi}\bigg[(1-\xi^2)\underbrace{\left(-\frac{1}{p}\frac{E}{E_{\rm c}}\bar{f}+\frac{1}{2}\nu_{\rm D}\frac{\partial \bar{f}}{\partial \xi}\right)}_{=0} -\underbrace{\frac{\xi(1\!-\!\xi^2)}{\taur \gamma}}_{\rm neglect}\bar{f}\bigg] \, \label{eq:FP2} 
\end{eqnarray}  
where $\bar{f} = p^2f$. 

Following the method and notation of~\cite{AleynikovPRL}, 
the condition that the pitch-angle flux vanishes yields the following
form for the angular distribution:  \begin{equation}
\bar{f} = G(t,p)A \exp(A
\xi)/2\sinh A, \label{eq:fbar}
\end{equation} where the parameter $A$ is defined as
\begin{equation*}
A(p) \equiv \frac{2 E}{p\nu_{\rm D} E_{\rm c}  }.
\end{equation*} 
 Then,~\eqref{eq:FP2}
integrated over pitch-angle yields a continuity equation
\begin{equation*}
\frac{\partial G}{\partial \tau} + \frac{\partial}{\partial p}
\left[U(p)G\right] =0,
\end{equation*}
where
\begin{eqnarray}
\fl U(p)=\frac{E/E_{\rm c}}{\tanh A} - \left[p\nu_{\rm s} +\Fbr +\frac{p \nu_{\rm D}}{2}
\right. \nonumber \\  + \left. 
%+
\frac{ p^2\gamma \nu_{\rm D}}{ \taur E/E_{\rm c} }\left(\frac{1}{\tanh A}-\frac{1}{A}\right)\right].
\label{eq:U}
\end{eqnarray}

As the sign of $U(p)$ determines  if the distribution at $p$ is accelerated or decelerated, the effective critical  electric field is the minimum electric  field for which force
balance is possible:
\begin{eqnarray}
E_{\rm c}^{\rm eff}&\equiv \min_p\left[E\big|U(p,E)=0\right] .
\label{eq:Eceffeq}
\end{eqnarray}

The minimum can be found analytically if $A\!\gg\! 1$ (so that $\tanh A \approx 1$) and the critical momentum fulfills 
$p_{\rm c}(\Eceffeq)\!\gg\! 1$, which are consistent with our final solution if partially ionized impurities dominate. Hence \eqref{eq:nuDApprox},~\eqref{eq:nuSApprox} and~\eqref{eq:phiBrems} may be used, and~\eqref{eq:Eceffeq} is approximately solved by (see~\ref{app:Derivation} for more details):~\footnote{A numerical implementation of \eqref{eq:EceffAnalytical} is available at \url{https://github.com/hesslow/Eceff}.}
\begin{eqnarray} 
\frac{\Eceffeq}{E_{\rm c}} \approx & \ \nuSD + \nuSH\!\left[\!\left(1\!+\!\frac{\nuDC}{\nuDB}\right)\ln\!\frac{\nuDB}{2\nuSH}+\sqrt{2\delta + 1}\right], \label{eq:EceffAnalytical}
\end{eqnarray}
where the constants are given in~\eqref{eq:lnLStar}, \eqref{eq:nuDB}, \eqref{eq:nuDC}, \eqref{eq:nuSD}, \eqref{eq:nuSH}, \eqref{eq:taurInv}, and \eqref{eq:phiBrems}, and
$\delta$, which is a measure of the effect of
radiation losses, is given by 
\begin{equation}
  \delta(\Eceffeq) =    \frac{ \nuDB}{ \nuSH^2}\left( \frac{\nuDB \taurInv}{ \Eceffeq/E_{\rm c}}+ \phibrI + \phibrII\ln (\nuDB/2\nuSH)\right) \label{eq:delta}.
\end{equation}
Since $\delta$ depends on \Eceff,~\eqref{eq:EceffAnalytical} is not in a closed form, and therefore~\eqref{eq:EceffAnalytical} and~\eqref{eq:delta} are evaluated iteratively starting at  $\Eceffeq = \Ectot$, where $\Ectot$ is the critical electric
field including the density of both bound and free electrons:
\begin{equation}
\Ectot \equiv \frac{n_{\rm e}^{\rm tot}}{n_{\rm e}}E_{\rm c}  = \frac{n_{\rm e}^{\rm tot} e^3 \lnLStar}{4 \pi \epsilon_0^2 m_{\rm e} c^2} \label{eq:Ectot},
\end{equation}
with $n_{\rm e}^{\rm tot} = n_{\rm e} + \sum_j n_j N_{{\rm e},j}$.
  Here, we iterate once so that $\delta_0 =
\delta(\Eceffeq\!=\!\Ectot)$ and $\delta \approx \delta_1 =
\delta[\Eceffeq(\delta_0)]$.
Equation~\eqref{eq:EceffAnalytical} was found to be accurate to within 10\% for magnetic fields in the range $B_{\rm T}^2\! \lesssim \! 100 \, n^{\rm tot}_{20}$ for all considered impurity species and plasma compositions. 

 \begin{figure}
 \centering
  \includegraphics[width=(0.3\linewidth+0.3\textwidth)]{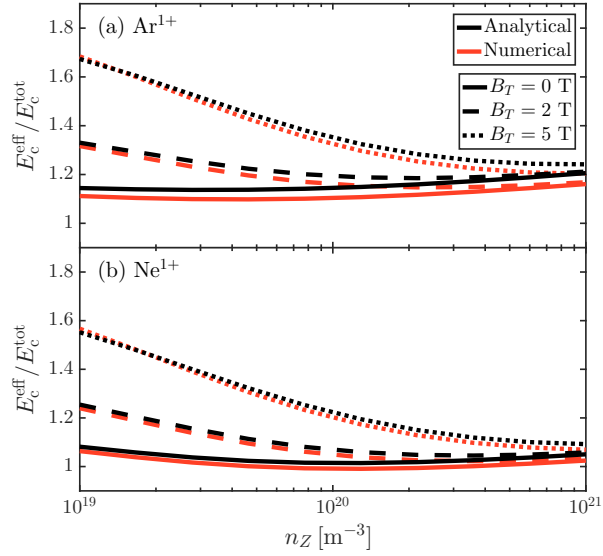}
 \caption{\label{Eceff}  Effective critical electric field normalized to $\Ectot$~\eqref{eq:Ectot} as
   function of $n_{Z}$, where $n_{ Z}$ is the
   density of  Ar$^{+}$ (top) and Ne$^{+}$ (bottom). The analytical expression~\eqref{eq:EceffAnalytical} is plotted in black, and the numerical solutions to~\eqref{eq:Eceffeq} are illustrated in red. The magnetic field is $B = \unit[0]{T}$ (solid line), $B = \unit[2]{T}$ (dashed line) and $B = \unit[5]{T}$ (dotted line).
   Parameters: $T = \unit[10]{eV}$, $n_{\rm D} =\unit[10^{20}]{m^{-3}} $. }
\end{figure}

Figure~\ref{Eceff} shows the effective critical electric field
normalized to $\Ectot$. Our model, corresponding to~\eqref{eq:EceffAnalytical}, is shown in black and compared to the full numerical solution to~\eqref{eq:Eceffeq} (using the algorithm in~\cite{fmincon}, implemented as \verb"fmincon" in \textsc{matlab}) for three different values of the magnetic field: $B=\unit[0]{T}$ in solid line, $B=\unit[2]{T}$ dashed and $B=\unit[5]{T}$ in dotted line.  These are shown for singly ionized argon in figure~\ref{Eceff}(a) and singly ionized neon in \ref{Eceff}(b). The behavior is only weakly dependent on ionization states; this is illustrated with neutral argon and Ar$^{4+}$ in figure~\ref{fig:EceffApp}. In addition, we find that the background deuterium density has a  negligible effect on \Eceff\ when $Zn_Z \gg n_{\rm D}$.

 \begin{figure}
 \centering
  \includegraphics[width=(0.3\linewidth+0.3\textwidth)]{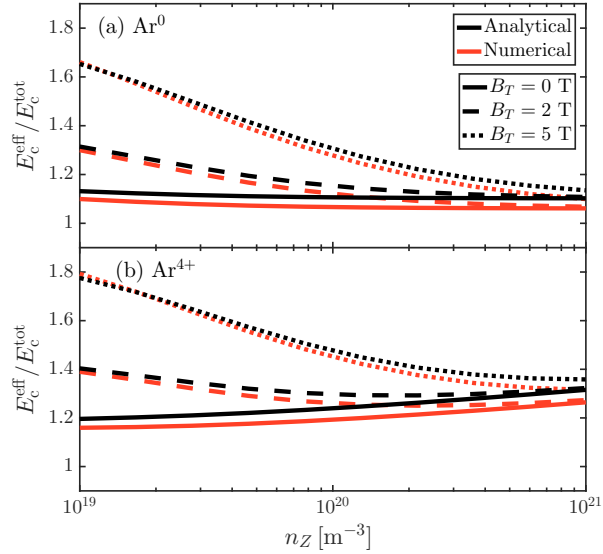}
 \caption{\label{fig:EceffApp} \small \noindent Effective critical electric field normalized to $\Ectot$~\eqref{eq:Ectot} as
   function of $n_{Z}$, where $n_{Z}$ is the
   density of (a) Ar$^{0}$ and (b) Ar$^{4+}$. The black lines correspond to the analytical expression~\eqref{eq:EceffAnalytical}, and the red lines are the numerical solutions to~\eqref{eq:Eceffeq}. The magnetic field is $B = \unit[0]{T}$ (solid line), $B = \unit[2]{T}$ (dashed line) and $B = \unit[5]{T}$ (dotted line).
   Parameters: $T = \unit[10]{eV}$, $n_{\rm D} =\unit[10^{20}]{m^{-3}} $.}
\end{figure}

Figures~\ref{Eceff}-\ref{fig:EceffApp} also show that with weakly ionized impurities, 
\begin{equation*}
\Eceffeq \gtrsim \Ectot \gg E_{\rm c}.
\end{equation*} 
Hence, it is more accurate to include \emph{all} electrons in the
critical electric field, than to count for instance half of the bound
electrons as done in the Rosenbluth-Putvinski model ($E_{\rm c}^{\rm
  RP}=\Ectot(n_{\rm e} + 0.5 n_{\rm bound})/n_{\rm e}^{\rm tot} $).  This
underestimation of the effective critical field by the RP model is a
result of using a
simplistic form of the inelastic collision rate as well as neglecting the effect of pitch-angle scattering and radiation losses.  To further explore the scaling of \Eceff\ with
magnetic field strength and impurity content, we approximate~\eqref{eq:EceffAnalytical} in the case where one weakly ionized
state $j$ dominates:
\begin{eqnarray}
\frac{\Eceffeq}{\Ectot}\! \approx \frac{n_{\rm e}}{n_{\rm e}^{\rm tot}}\!+\! \frac{ N_{{\rm e},j} n_Z}{n_{\rm e}^{\rm tot}} 
\frac{1}{\lnLStar}\!\!
%\nonumber\\ &\times\!
 \left(
 \! S_j\! +\! 
 R_j \sqrt{\frac{B_{\rm T}^2}{n_{20}^{\rm tot }} \! +\! \frac{0.9}{\ln \aj}}
\!\right)\!,\!\label{eq:simpleEceff} \\
 S_j  = \left[ \ln I_j^{-1}\!-\!1\!+\!\frac{3}{2}\left(\!1 \!+\! \frac{1}{\ln \aj
 }\right)
 \ln\!\left(\!\frac{Z_j}{3}   \ln \aj \right)\right],\! \label{eq:Sj}\\
 R_j = 0.09 (Z\!+\!Z_0)\ln \aj.
\end{eqnarray}
The screening constant $S_j$ is given for all argon and neon species in table~\ref{tab:constants} in~\ref{app:const}. For typical magnetic fields, the terms inside the brackets tend to be roughly 1-2 times $\lnLStar$. As $n_{\rm e}\! +\! N_{{\rm e},j} n_j\! =\! n_{\rm e}^{\rm tot}$ with only one impurity species $j$, one obtains $\Eceffeq \!\gtrsim \!\Ectot$. 
From~\eqref{eq:simpleEceff}, we thus conclude that the effect of partially stripped impurities scale approximately linearly with impurity density; more specifically, $\Eceffeq=\Ectot(n_{\rm e} + \kappa n_{\rm bound})/n_{\rm e}^{\rm tot}\approx \kappa \Ectot $, where $\kappa$ is between 1 and 2. Consequently, our calculated of \Eceff\ is up to  $4 E_{\rm c}^{\rm
  RP}$ in typical tokamak scenarios.

The radiation term $R_j$ quantifies the effect of bremsstrahlung and synchrotron losses; these are dominated by synchrotron radiation reaction if
\begin{equation*} B_{\rm T}^2 \gtrsim 0.2 n_{20}^{\rm tot}\,,
\end{equation*}
which is lower than the fully ionized estimation~\eqref{eq:synchbrems}. 
In this case, \Eceff\ depends linearly on $B_{\rm T}/\sqrt{n_{20}^{\rm tot }}$ . This agrees with the scaling found in~\cite{AleynikovPRL} for the fully ionized case. In contrast, for low magnetic fields, bremsstrahlung can increase the effective critical field by up to 20\% for argon. This number is insensitive to the plasma density and depends only on its ionic composition.

\section{Current decay}
\label{sec:decay}

The critical electric field, especially as modified by the effects of partially screened nuclei and radiation losses, plays an important role during the relaxation of runaway electrons. In this section, we demonstrate with kinetic simulations that~\eqref{eq:EceffAnalytical} well characterizes the threshold between runaway growth and decay under these modifications. Then, when the electric field evolves self-consistently, we show that it remains tied to $E_{\rm c}^{\rm eff}$ under certain assumptions during the current decay phase of a tokamak disruption.

If the current is carried by runaway electrons and the shape of the runaway distribution is constant in time,
the time derivative of the current is related to the steady-state runaway growth rate 
\begin{equation}
   \Gamma(E) \equiv \frac{1}{n_{\rm RE}}\frac{\mathrm{d} n_{\rm RE}}{\mathrm{d}t} \approx  \frac{1}{I} \frac{\mathrm{d} I}{\mathrm{d}t}.
\end{equation}
The scaling of the growth rate with impurity content may be estimated from the  Rosenbluth--Putvinski formula~\cite{rosenbluthPutvinski}  by replacing $E_{\rm c}$ with \Eceff\ and the density by the total electron density  due to the fact that bound and free electrons have equal probability of
becoming runaway electrons through knock-on collisions:
\begin{equation}
\Gamma(E)  \sim  
\frac{1}{\lnLStar}\frac{1}{\tau_{\rm c}^{\rm tot}}\left(\frac{E}{\Eceffeq}-1\right),
\label{eq:Gamma}
\end{equation}
with  
$\tau_{\rm c}^{\rm tot} \!= \!(\nofrac{n_{\rm e}}{n_{\rm e}^{\rm
    tot}})\tau_{\rm c}$.
The qualitative scaling of the analytic growth rate is confirmed in
figure~\ref{fig:Gamma}, where the growth rate is numerically
calculated using \textsc{code}~\cite{CODEPaper2014,Stahl2016}, which directly 
solves the kinetic equation~\eqref{eq:FP}.
These simulations employed the general
field-particle knock-on operator of~\cite{olaknockon2018,CQL1,CQL2} and a Boltzmann operator for
partially screened bremsstrahlung losses as described in section~\ref{subsec:rad}.  
The vertical lines denote the analytic prediction in~\eqref{eq:EceffAnalytical} for when
one would expect the transition between growth and decay of an existing 
runaway population. Radiation losses affect where this threshold lies 
and the analytic model $\Eceffeq$ accurately and robustly captures 
this effect. In particular, we note that the mean-force bremsstrahlung 
model employed in the analytical derivation
of \Eceff\ agrees with the Boltzmann-type bremsstrahlung operator used
in the simulations within a few percent. 

 \begin{figure}
 \centering
  \includegraphics[width=(0.3\linewidth+0.3\textwidth)]{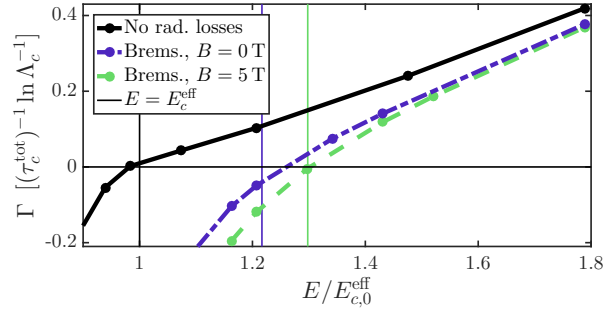}
 \caption{\label{fig:Gamma} \small \noindent Steady-state runaway
   growth rate as a function of electric field normalized to the
   critical electric field $E_{c,0}^{\rm eff}$ without radiation
   losses.  The solid black line is without radiation losses; the
   dash-dotted blue line includes bremsstrahlung and the dashed green
   line includes both bremsstrahlung and synchrotron losses
   corresponding to $\unit[B=5]{T}$.  The vertical lines denote the
   analytical prediction $E =\Eceffeq $.  Parameters: $n_{\rm
     D}=\unit[10^{20}]{m^{-3}}$, a density of Ar$^{+}$ given by
   $n_{\rm Ar}=4 n_{\rm D}$ and $T =
   \unit[10]{eV}$. }
\end{figure}

The electric field is hypothesized to remain close to $\Eceffeq$ during 
the current-decay phase of a tokamak disruption~\cite{breizman2014}. 
The mechanism by which this occurs is the fast timescale of the avalanche generation in relation to the inductive timescale of the system. 
A toroidal electric field is induced when there is a time-changing magnetic flux 
through a current loop such as a runaway beam. This magnetic flux is proportional
to the total current through the loop. The induced electric field is therefore
related to the rate of change of the current:
\begin{equation}
E  =-\frac{L}{2 \pi R}\frac{{\rm d} I}{{\rm d} t}\label{eq:ind},
\end{equation}
where $R$ is the major radius of the tokamak. This inductance model has recently been implemented in \textsc{code} to calculate the electric field self-consistently with the evolution of the electron velocity distribution.
In general, the exact value of the inductance $L$ will depend on the spatial distribution of 
current, which will change in time. For a large-aspect ratio current loop (such as a runaway beam),  $L$ can be approximated by~\cite{inductance}
\begin{equation} \label{eq:loopinductance}
   L \approx \mu_0 R \left[ \ln \left( \frac{8 R}{a} \right)-2+ \frac{l_i}{2} \right].
\end{equation} 
Here, $R$ is the major radius of the tokamak, $a$ is the radius of the runaway beam, and $l_i$ parametrizes the distribution of current within the beam. 
%In experimentally relevant scenarios, the value of $L$ only changes $\lesssim 10\%$ depending on the current distribution. 
We have chosen $l_i = 1.5$ as a representative mid-plateau value, based on experimental results from European medium sized tokamaks.

%Equilibrium reconstructions show that $l_i\approx 1$ during the ramp-up phase, and slowly increases to $l_i \approx 1.5$ as the beam evolves. For the simulations shown below, the choice results in a difference of about 5-13\% in the value of $L$. Since the current profile is not evolved self-consistently in these 0-D simulations, we choose the constant value $l_i = 1.5$ when estimating the inductance with Eq.~\eqref{eq:loopinductance}. 

When $E\approx \Eceffeq$,  the growth rate can be expanded according to
\begin{equation*} 
\Gamma = \Gamma'\!(\!\Eceffeq\!) [E-\Eceffeq] +\dots, 
 \end{equation*}
which allows~\eqref{eq:ind} to be solved analytically:
\begin{eqnarray}
   E \approx \Eceffeq\left(1 - \frac{2 \pi R}{L I_{\rm RE} \Gamma'\!(\!\Eceffeq\!) } \right) .
   \label{eq:inducedE}
\end{eqnarray}
This yields a condition under which the electric field remains close to $\Eceffeq$:
\begin{equation*}L I_{\rm RE} \Gamma'\!(\!\Eceffeq\!) \gg 2 \pi R. 
\end{equation*}
With the estimate of $ \Gamma'\!(\! E\!)$ from the numerical 
results of figure~\ref{fig:Gamma} (at $\unit[B=0]{T}$) and estimating $R/a \! \approx\! 5$
we find that the minimum required current for
$E\!\approx\! \Eceffeq$ is approximately
\begin{equation}
I_{\rm RE}\, \gg \unit[60]{kA}.
\label{eq:MLarge}
\end{equation}
This value is substantially lower than the estimation of \unit[250]{kA} in~\cite{breizman2014}, which did not include the effect of
partial screening or radiation losses. Since this threshold current is inversely proportional to the inductance, the estimate~\eqref{eq:MLarge} is only weakly dependent on the details of the spatial current distribution. Therefore, the exact value of the instantaneous inductance does not affect the primary result of this section: for large enough inductance, the electric field remains approximately tied to $\Eceffeq$ during the current decay phase, leading to a predictable   decay time scale.

To test the hypothesis that $E\!\approx\! \Eceffeq$ when $I_{\rm RE}\! \gg\! \unit[60]{kA}$, we generate a forward-beamed initial distribution obtained from a
simulation with a large electric field; the initial average runaway
energy in our simulation is \unit[17.2]{MeV}. We then inject singly ionized
argon with a density that is four times the deuterium density
$n_\mathrm{D}\!=\!\unit[10^{20}]{m^{-3}}$. Starting at an initial current
density $j_0 \!= \!\unit[12.9]{MA /m^{2}}$, we let the electron
distribution evolve with a self-consistent electric field in a
strongly, intermediate or weakly inductive system. At a constant
current density, varying $I_0^{\rm RE} L/(\mu_0 R)$ corresponds to varying
$L/(\mu_0 R) $ through the beam aspect ratio $R/a$ or the initial current
$I_0\!= \!j_0\pi a^2 $. The following values were chosen in the simulations:
$\pi a^2 L/(\mu_0 R) \! =\!  (i)\, 4.30\,, (ii)\, 1.57\,{\rm and}\,
(iii)\, 0.14$. If $R/a\!=\!5$ and $l_i=1.5$, these three values correspond
to an initial current of $(i)\, I_0^{\rm RE}\!=\!\unit[23]{MA}$; $(ii)\, I_0^{\rm RE}\!=\!\unit[8.3]{MA}$; and $(iii)\,I_0^{\rm RE}\!=\!\unit[0.75]{MA}$. 
   As in the growth rate
simulations, we include both synchrotron losses, the full
bremsstrahlung model and a Chiu-Harvey type avalanche operator.

Figure \ref{decay}a shows the current decay, which is linear (as
expected) and faster in the low inductance case.  Figure \ref{decay}b
shows the electric field evolution. Clearly, in the high-inductance
case, the electric field is close to the critical field after an
initial transient. 
 This means that, in highly inductive devices such
as ITER, the current decay is to a very good approximation given by
${\rm d} I_{\rm RE} /{\rm d} t= -2 \pi R E_{\rm c}^{\rm eff} /L$. Enhanced $E_{\rm c}^{\rm eff}$ will lead to faster current decay,
and~(\ref{eq:EceffAnalytical}) quantifies how fast the decay
is.

On the other hand, the induced electric field deviates by approximately 10\% from \Eceff\ in the low-inductance case. Since the initial current $I_0^{\rm RE}\!= \!\unit[750]{kA}$ is high in relation to many medium-sized tokamak experiments, $E\!\approx \!\Eceffeq$ gives an overestimation of the current decay rate in many of today's devices. The relative deviation from \Eceff\ observed in figure~\ref{decay}b is consistent with the estimation $1-E/\Eceffeq \approx \unit[60]{kA}/I_{\rm RE}$ from~\eqref{eq:inducedE} and \eqref{eq:MLarge}. 

Although the predicted induced electric field obeys $E\!\leq\! \Eceffeq$ with our assumptions, several effects could lead to a  higher induced electric field in an actual experimental discharge. For example, a stronger electric field would be necessary to balance a runaway population with sub-relativistic energy, in which case the steady-state growth rate used here is inaccurate. Other effects such as transport~\cite{Zeng2013,Papp2015,Ficker2017}, trapping~\cite{rosenbluthPutvinski,Nilsson2015} and wave-particle interaction~\cite{BA2017,FN2014,Pokol2014,LiuWhistler2018} may also increase the runaway current decay rate and accordingly the induced electric field. Such complete modelling remains the subject of future work. Nevertheless, partial screening has a major effect on the critical electric field as demonstrated here, and therefore the results derived herein should be an important piece toward improved experimental comparison of the runaway current decay rate as well as the avalanche growth rate.

 \begin{figure}
 \centering
  \includegraphics[width=(0.3\linewidth+0.3\textwidth)]{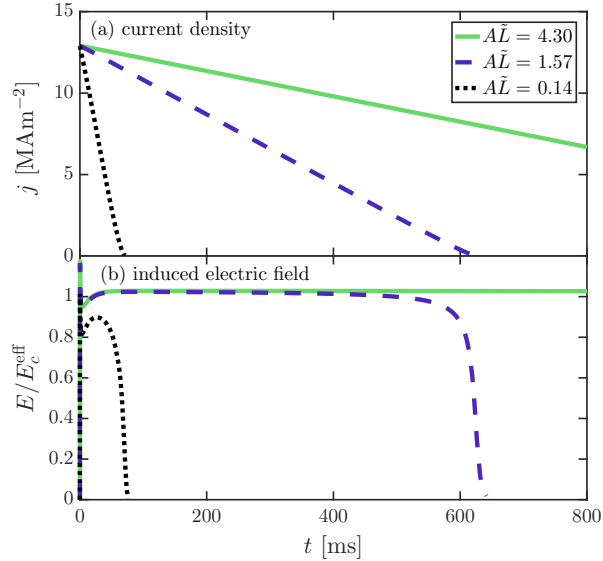}
 \caption{\label{decay} \small \noindent Current decay (top) and
   electric field (bottom) for $T = 10 \; \rm eV$, Ar$^{+}$ with
   $n_\mathrm{Ar}\! =\! 4 n_\mathrm{D}$, $n_\mathrm{D}\!=\!10^{20}$
   m$^{-3}$, for three different inductance parameters $A \tilde{L} \equiv \pi a^2 L/(\mu_0 R)$ in solid blue, dashed green and dotted black line respectively. The initial
   average runaway energy was 17.2 MeV. Bremsstrahlung losses were included here, and $\unit[B=0]{T}$ for simplicity.}
\end{figure}

Finally, we note that the simulations with an inductive electric field
validate the initial assumption of rapid pitch-angle dynamics in~\eqref{eq:FP2}; we find that the resulting pitch-angle
distribution in~\eqref{eq:fbar} is accurate for
$E\!\approx \!\Eceffeq$; see~\ref{app:dist}. 
The distribution function in~\eqref{eq:fbar} is
consequently appropriate for determining the effective critical
electric field, but not for describing runaway generation.

\section{Conclusion}
\label{sec:concl}
Recent experimental studies on several tokamaks show that the onset
and decay of runaway electrons occurs for critical electric fields
that are considerably higher than the Connor--Hastie field $E_{\rm c}$. One reason is that there are
other runaway loss mechanisms in addition to damping due to collisions
in a fully ionized plasma that seem to dominate both in disruptive and
quiescent cases. In this paper, we show that if there are heavy
partially ionized impurities present in the plasma, the dominant
effect on the critical electric field is the effect of partial
screening.  The effective critical field is further increased due to
the enhanced radiation loss rates when partially ionized impurities
are present.

We give analytical formulas for the effective critical electric field \Eceff\
including partial screening and radiation effects, derived under the
condition of rapid pitch-angle dynamics.  The validity of this
assumption and the value of the effective critical electric field is
demonstrated by numerical simulations with the kinetic equation solver
\textsc{code}. The most complete expression for the critical electric field is
given in~\eqref{eq:EceffAnalytical}. It has been shown to be valid for a
wide range of magnetic fields, impurity species and plasma
composition. 
To make the parametric dependencies more transparent, 
we also give an approximate expression
in~\eqref{eq:simpleEceff} that is valid when one weakly ionized
state dominates, which is often the case in a cold post-disruption
tokamak plasma.

As expected, we find that in the presence of large amounts of heavy
impurities, the effective critical field can be drastically higher
than  $E_{\rm c}$ which is proportional to the
density of free electrons: \Eceff\ even exceeds the value obtained by
including the total density of both free and bound electrons. 
In contrast to Rosenbluth--Putvinski~\cite{rosenbluthPutvinski}, where the 
effective density includes half of the bound electrons, $n = n_{\rm e} + 0.5 n_{\rm bound}$, our calculations show that the bound electrons are  weighted by a factor of typically 1-2. This enhancement is attributed to the energy-dependent collisional friction, pitch-angle scattering as well as radiation losses. Bremsstahlung
and synchrotron losses both increase the effective critical field,
typically by tens of percent.

Using a 0D inductive electric field we calculate the runaway current
decay after impurity injection. 
Through kinetic simulations we confirm the accuracy of the formula for the effective critical
field~\eqref{eq:EceffAnalytical}, and demonstrate that the electric field stays 
close to the effective critical field when the runaway current satisfies $I_{\rm RE}\! \gg \! \unit[60]{kA}$, in which case $ {\rm d} I_{\rm RE} /{\rm d} t\propto\Eceffeq$. These findings are relevant for
the efficacy of mitigation strategies for runaway electrons in tokamak
devices: since the runaway current decay rate is typically 2-4 times higher than what is predicted by the Rosenbluth--Putvinski formula, a lower quantity of assimilated material is required for successful mitigation.
As screening significantly increases the critical electric field, we anticipate that this effect is of importance to include in experimental comparisons;  however, accurate predictions may require the modelling of spatial effects which are not considered here. 

\ack
The authors are grateful to E Hollmann, S Newton and A Stahl for stimulating discussions and to T DuBois and M Rahm for the simulations needed to determine the effective ion size.  This work was supported by the   Swedish Research Council (Dnr.~2014-5510), the Knut and Alice   Wallenberg Foundation and the European Research Council (ERC-2014-CoG grant 647121). The work has been carried out within the framework of the EUROfusion Consortium and has received funding from the Euratom research and training programme 2014-2018 under grant agreement No 633053. The views and opinions expressed herein do not necessarily reflect those of the European Commission.
\appendix

\section{Constants for the effective electric field}
\label{app:const}
\Tref{tab:constants} summarizes the constants needed to compute the value of the
effective electric field in the presence of argon and neon. The
effective ion size $\aj$ is determined by DFT simulations
and is related to $a_j$ in~\cite{Hesslow} through $\aj = 2 a_j/\alpha $ where $\alpha \approx 1/137$ is the fine-structure constant.
The mean
excitation energy $I_j$ is taken from~\cite{sauer2015}.  These
give $S_j$ from~\eqref{eq:Sj} according to
\begin{equation*} 
S_j  =  \left[ \ln I_j^{-1}\!-\!1\!+\!\frac{3}{2}\left(\!1 \!+\! \frac{1}{\ln \aj
 }\right)
 \ln\!\left(\!\frac{Z_j}{3}   \ln \aj \right)\right].
 \end{equation*}

\begin{table*}[htbp]
\caption{\label{tab:constants} Constants to determine \Eceff.}
\begin{indented}
\lineup
\item[]
\begin{tabular}{@{}lllllllll}
\br
&$\ln\aj$&$\ln \IjInv$& $S_j$&\phantom{.}&&\textbf{$\ln\aj\!$}&\textbf{$\ln \IjInv$}& $S_j$ \\ \cline{1-4}\cline{6-9}\tallrow
Ar$^{ 0}$&4.6&7.9&13.0&&Ne$^{0}$&4.7&8.2&12.2 \\ 
\tallrow
Ar$^{ 1+}$&4.5&7.8&12.8&&Ne$^{1+}$&4.6&8.0&12.0 \\ \tallrow
Ar$^{ 2+}$&4.4&7.6&12.6&&Ne$^{2+}$&4.5&7.9&11.8 \\ \tallrow
Ar$^{ 3+}$&4.4&7.5&12.5&&Ne$^{3+}$&4.4&7.7&11.6 \\ \tallrow
Ar$^{ 4+}$&4.3&7.3&12.3&&Ne$^{4+}$&4.3&7.5&11.4 \\ \tallrow
Ar$^{ 5+}$&4.2&7.2&12.2&&Ne$^{5+}$&4.1&7.3&11.2 \\ \tallrow
Ar$^{ 6+}$&4.1&7.0&12.0&&Ne$^{6+}$&4.0&7.0&10.8 \\ \tallrow
Ar$^{ 7+}$&4.0&6.8&11.8&&Ne$^{7+}$&3.7&6.6&10.4 \\ \tallrow
Ar$^{ 8+}$&3.9&6.6&11.5&&Ne$^{8+}$&3.2&5.9&\09.5 \\ \tallrow
Ar$^{ 9+}$&3.8&6.5&11.4&&Ne$^{9+}$&3.1&5.8&\09.5 \\ \tallrow
Ar$^{10+}$&3.7&6.4&11.3&& & & &  \\ \tallrow
Ar$^{11+}$&3.6&6.2&11.1&& & & &  \\ \tallrow
Ar$^{12+}$&3.6&6.1&11.0&& & & &  \\ \tallrow
Ar$^{13+}$&3.5&5.9&10.8&& & & &  \\ \tallrow
Ar$^{14+}$&3.3&5.7&10.5&& & & &  \\ \tallrow
Ar$^{15+}$&3.1&5.3&10.1&& & & &  \\ \tallrow
Ar$^{16+}$&2.6&4.7&\09.4&& & & &  \\ \tallrow
Ar$^{17+}$&2.5&4.7&\09.4&& & & & 
\\ \br
\end{tabular}

% Alternative form of the table with half the width
%\begin{tabular}{@{}llll}
%\br 
%&$\ln\aj$&$\ln \IjInv$& $S_j$ \\ \mr \tallrow
%Ar$^{ 0}$&4.6&7.9&13.0 \\ \tallrow
%Ar$^{ 1+}$&4.5&7.8&12.8 \\ \tallrow
%Ar$^{ 2+}$&4.4&7.6&12.6 \\ \tallrow
%Ar$^{ 3+}$&4.4&7.5&12.5 \\ \tallrow
%Ar$^{ 4+}$&4.3&7.3&12.3 \\ \tallrow
%Ar$^{ 5+}$&4.2&7.2&12.2 \\ \tallrow
%Ar$^{ 6+}$&4.1&7.0&12.0 \\ \tallrow
%Ar$^{ 7+}$&4.0&6.8&11.8 \\ \tallrow
%Ar$^{ 8+}$&3.9&6.6&11.5 \\ \tallrow
%Ar$^{ 9+}$&3.8&6.5&11.4 \\ \tallrow
%Ar$^{10+}$&3.7&6.4&11.3 \\ \tallrow
%Ar$^{11+}$&3.6&6.2&11.1  \\ \tallrow
%Ar$^{12+}$&3.6&6.1&11.0 \\ \tallrow
%Ar$^{13+}$&3.5&5.9&10.8 \\ \tallrow
%Ar$^{14+}$&3.3&5.7&10.5 \\ \tallrow
%Ar$^{15+}$&3.1&5.3&10.1 \\ \tallrow
%Ar$^{16+}$&2.6&4.7&\09.4 \\ \tallrow
%Ar$^{17+}$&2.5&4.7&\09.4 \\ \tallrow
%Ne$^{0}$&4.7&8.2&12.2 \\ \tallrow
%Ne$^{1+}$&4.6&8.0&12.0 \\ \tallrow
%Ne$^{2+}$&4.5&7.9&11.8 \\ \tallrow
%Ne$^{3+}$&4.4&7.7&11.6 \\ \tallrow
%Ne$^{4+}$&4.3&7.5&11.4 \\ \tallrow
%Ne$^{5+}$&4.1&7.3&11.2 \\ \tallrow
%Ne$^{6+}$&4.0&7.0&10.8 \\ \tallrow
%Ne$^{7+}$&3.7&6.6&10.4 \\ \tallrow
%Ne$^{8+}$&3.2&5.9&\09.5 \\ \tallrow
%Ne$^{9+}$&3.1&5.8&\09.5 
%\\ \br
%\end{tabular}

\end{indented}
\end{table*} 

\section{Angular dependence of the runaway electron distribution function}
\label{app:dist}

The simulations with an inductive electric field (figure~\ref{decay})
can be used to validate the initial assumption of rapid pitch-angle
dynamics in~\eqref{eq:FP2} leading to the pitch-angle distribution
in~\eqref{eq:fbar}.
Expanding $\bar f$ in Legendre polynomials
\begin{equation*}
\bar f = \sum_L \bar f_L(p) P_L(\xi),
\end{equation*} we relate the predicted analytical distribution in~\eqref{eq:fbar} to the ratio between the zeroth and the first Legendre modes of the distribution:  
\begin{equation}
\frac{1}{3}\frac{\bar f_1}{\bar f_0} = \left(\frac{1}{\tanh
  A}-\frac{1}{A}\right)
\label{eq:predictedRatio}.
\end{equation}
The ratio given in~\eqref{eq:predictedRatio} quantifies the
narrowness of the electron distribution: $\nofrac{\bar f_1}{3\bar
  f_0}=0$ corresponds to an isotropic distribution while the
$\nofrac{\bar f_1}{3\bar f_0}\rightarrow 1$ for a narrow, beam-like
distribution.  Figure~\ref{fig:A} compares the numerical value of
$\nofrac{\bar f_1}{3\bar f_0}$ as computed in \textsc{code} in solid black
line, to the analytical prediction~\eqref{eq:predictedRatio} in
dashed green line.  The analytical formula accurately predicts the
distribution width on the entire interval from a fully isotropic
distribution at $p=0$ to a narrow beam for $p\gg1$. This validates our
assumptions on the rapid pitch-angle dynamics in~\eqref{eq:FP2}.  In
contrast, for larger electric fields ($E/\Eceffeq \gtrsim 5$), we find
that the distribution rather follows the formula in
F\"ul\"op et al.~\cite{Fulop2006}, which is derived in the limit of $E\gg
\Eceffeq$. 

 \begin{figure}
 \centering
  \includegraphics[width=(0.3\linewidth+0.3\textwidth)]{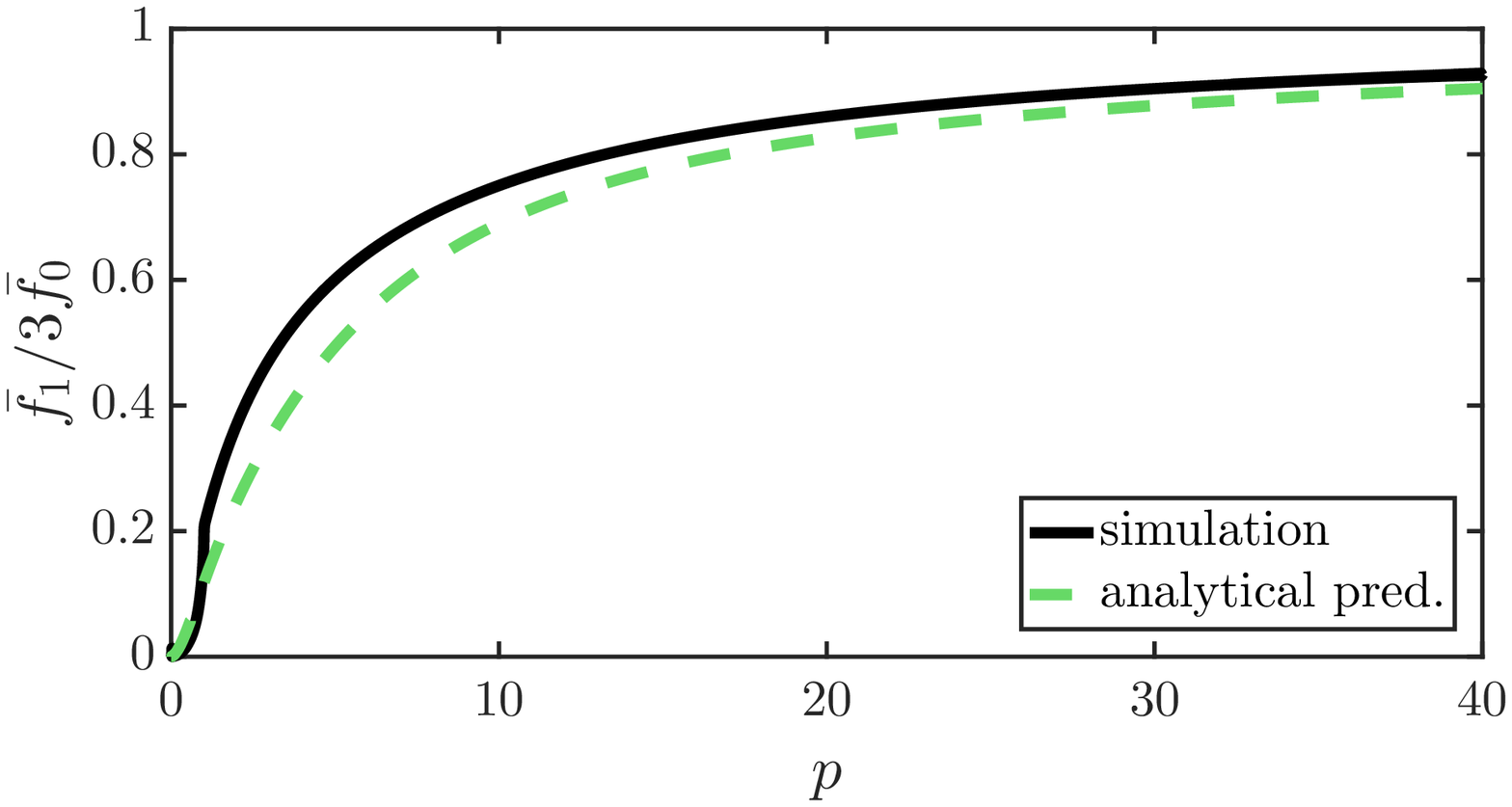}
 \caption{\label{fig:A} \small \noindent 
 The distribution width parameter $\nofrac{\bar f_1}{3\bar f_0}$ as a function of momentum $p$ taken after \unit[200]{ms} for the high inductance case in figure~\ref{decay}. This snapshot is representative for all times and for both the intermediate and the high-inductance cases. }
\end{figure} 

\section{Derivation of the effective critical field}
\label{app:Derivation}
The effective critical field can be found analytically 
noting that the critical momentum % -- which is finite  because of partial screening, radiation reaction losses as well as an energy-dependent Coulomb logarithm -- 
fulfills $\pcStar \equiv p_{\rm c}(\Eceffeq)\!\gg\! 1$. 
Moreover, we assume that $A$, which is defined in~\eqref{eq:fbar}, fulfills $A\!\gg\! 1$ (so that $\tanh A \approx 1$). These two assumptions are consistent with our final solution if partially ionized impurities dominate. 
Hence \eqref{eq:nuDApprox},~\eqref{eq:nuSApprox} and~\eqref{eq:phiBrems} may be used in the expression for the effective critical field~\eqref{eq:Eceffeq}, and the requirement $U(p) = 0$ [with $U$ given in~\eqref{eq:U}] results in a quadratic equation in $E/E_{\rm c}$: 
%\begin{equation*}
% \frac{E}{E_{\rm c}} - \left[p\nu_{\rm s} +\Fbr +\frac{p \nu_{\rm D}}{2}
%+
%\frac{ p^2\gamma \nu_{\rm D}}{ E \taur}\right]=0,
%\end{equation*}
%which can be rewritten as
\begin{equation}
\left(\frac{E}{E_{\rm c}}\right)^2 - \frac{E}{E_{\rm c}} h(p) -\epsilon(p)=0, \label{eq:quadratic}
\end{equation}
where $h(p)$ and $\epsilon(p)$ are both positive functions of $p$ within the assumption $p_{\rm c}\!\gg\! 1$:
\begin{eqnarray*}
\fl h(p) \equiv \nuSD + \nuSH \ln p + \frac{1}{2 p} (\nuDB + \nuDC \ln p)  \\+ p (\phibrI + \phibrII\ln p),\\
\fl \epsilon(p) \equiv  p(\nuDB + \nuDC \ln p)\taurInv.
\end{eqnarray*}
Consequently, finding the effective critical field amounts to evaluating the positive solution to~\eqref{eq:quadratic}
\begin{equation}
\frac{\Eceffeq}{E_{\rm c}} \approx \frac{1}{2}\left[h(\pcStar) + \sqrt{h(\pcStar)^2+4\epsilon(\pcStar)} \right]
\label{eq:EvaluateEceff}
\end{equation}
at the minimum $ p_{\rm c}^\star$, the critical momentum which minimizes $\Eceffeq$ in~\eqref{eq:quadratic}, which is determined by 
\begin{equation}
\frac{\Eceffeq}{E_{\rm c}} h'(p_{\rm c}^\star) +\epsilon'(p_{\rm c}^\star)=0.
\label{eq:definePc}
\end{equation}
The derivatives of $h(p)$ and $\epsilon(p)$ are given by
\begin{eqnarray*}
h'(\pcStar) &\approx \frac{\nuSH}{\pcStar}-\frac{1}{2 (\pcStar)^2}\left[\nuDB +\nuDC( \ln \pcStar -1 ) \right]  + \phibrI + \phibrII(\ln \pcStar + 1), \\
\epsilon'(\pcStar) &\approx  \taurInv[\nuDB + \nuDC(\ln \pcStar + 1)],
\end{eqnarray*}
and thus~\eqref{eq:definePc} is solved by
\begin{equation*}
\pcStar \approx \frac{2 \pcStarO}{1+\sqrt{1+2\delta}}\,, 
\end{equation*}
 where
\begin{eqnarray}
 \delta  \equiv  \frac{ \nuDB +\nuDC( \ln \pcStar\! -\!1 ) }{ \nuSH^2}[\phibrI + \phibrII(\ln \pcStar + 1)]( \x+  1), \label{eq:deltafirst}\\
 \pcStarO   \equiv  \frac{\nuDB +\nuDC( \ln \pcStar\! -\!1 ) }{2\nuSH} ,
 \\
\x \equiv  \frac{[\nuDB + \nuDC(\ln \pcStar + 1)] \taurInv}{[\phibrI + \phibrII(\ln \pcStar + 1)] \Eceffeq/E_{\rm c}}.
\end{eqnarray}
Here, $\x$ describes the relative importance of  synchrotron radiation compared to bremsstrahlung. 

 To evaluate~\eqref{eq:EvaluateEceff}, we first simplify $h(\pcStar)$ using $(1+\sqrt{1+2\delta})^{-1} = (\sqrt{1+ 2\delta}-1)/2 \delta$% and $\pcStarO/\delta = \nuSH/\{2(\x+1)[\phibrI + \phibrII(\ln \pcStar + 1)]\}$
 :
 \begin{eqnarray}
\fl h(\pcStar) = \nuSD  + \nuSH \ln \pcStar + \frac{\nuSH}{2}\frac{ (\nuDB + \nuDC \ln \pcStar) }{\nuDB +\nuDC( \ln \pcStar -1 ) }\big(1+\sqrt{2 \delta +1 }\big) \nonumber\\
+\frac{\nuSH}{2} \frac{\phibrI + \phibrII\ln \pcStar}{\phibrI + \phibrII(\ln \pcStar + 1)} \frac{\sqrt{1+2\delta}-1}{\x + 1} \nonumber\\
\fl \qquad \approx
\underbrace{\nuSD  + \nuSH \left(\ln \pcStar + 1 + \frac{\nuDC}{\nuDB} \ln \pcStar\right)}_{\equiv h_0(\pcStar)} \nonumber\\ + 
\frac{\nuSH}{2} \big(\sqrt{1+2\delta}-1 \big) \left(\frac{\x+2 }{\x +1} \right),
\label{eq:part1}
\end{eqnarray}
where we assumed $\nuDB \gg \nuDC( \ln \pcStar -1 ) $ since $\nuDB \gg \nuDC$ typically; see \eqref{eq:nuDB} and~\eqref{eq:nuDC}. Furthermore,  we assumed $\phibrII \ll \phibrI + \phibrII \ln \pcStar $.
To simplify $\epsilon(\pcStar)$, we approximate $\Eceffeq/E_{\rm c} \approx  h(\pcStar)$ and assume $\nuDC \ll  \nuDB + \nuDC \ln \pcStar $:
\begin{eqnarray}
 \epsilon(\pcStar) &= \frac{\nuSH}{2} \Eceffeq \big(\sqrt{1+2\delta}-1 \big) \frac{\nuDB + \nuDC \ln \pcStar}{\nuDB + \nuDC(\ln \pcStar + 1)}\left(\frac{\x}{ \x + 1} \right)
\nonumber\\ &\approx
\frac{\nuSH}{2} h(\pcStar) \big(\sqrt{1+2\delta}-1 \big) \left(\frac{\x}{ \x + 1} \right).
 \label{eq:estEpsilon}
\end{eqnarray}
Then, 
\begin{eqnarray}
\fl \sqrt{h(\pcStar)^2+4\epsilon(\pcStar)}&\approx
 \sqrt{h_0(\pcStar) + \frac{\nuSH}{2}\big(\sqrt{1+2\delta}-1 \big)
 \frac{\x +2 }{\x +1} }  \nonumber \\ &\quad \times \sqrt{h_0(\pcStar) + \frac{\nuSH}{2}\big(\sqrt{1+2\delta}-1 \big)
 \frac{5\x +2 }{\x +1} } \nonumber \\
&\approx \left(h_0(\pcStar) +\frac{\nuSH}{2} \big(\sqrt{1+2\delta}-1 \big)
 \frac{3\x +2 }{\x +1} \right), 
 \label{eq:part2}
\end{eqnarray}
where the last approximation is a matching between the behavior at $ \x \gg 1$ and $\x \ll 1$ for $2 h_0(\pcStar) \gg \nuSH(\sqrt{1+2\delta} -1)$, i.e.\ screening effects dominate over radiation reaction effects. This assumption also motivates the approximation 
\begin{equation}
 \ln \pcStar \approx
\ln \pcStarO \approx \ln (\nuDB/2\nuSH) \label{eq:pcstar}.
\end{equation}
Finally, the effective critical field~\eqref{eq:EvaluateEceff} is the mean of~\eqref{eq:part1} and~\eqref{eq:part2}: 
\begin{eqnarray} 
\frac{\Eceffeq}{E_{\rm c}} &\approx & 
h_0(\pcStar) + \nuSH\left(\sqrt{1+2\delta}-1 \right)  \nonumber \\
&\approx & \nuSD + \nuSH\!\left[\!\left(1\!+\!\frac{\nuDC}{\nuDB}\right)\ln\!\frac{\nuDB}{2\nuSH}+\sqrt{2\delta + 1}\right]. \label{eq:EceffAnalyticalApp}
\end{eqnarray}
For $\delta$ in equation~\eqref{eq:deltafirst}, we again approximate $\ln \pcStar$ using~ \eqref{eq:pcstar}  but also neglect the $\nuDC$ terms compared to $\nuDB$, which is motivated both by the smallness of $\nuDC$ compared to $\nuDB$ and the fact that~\eqref{eq:pcstar}  overestimates $\ln \pcStar$ if the effect of radiation reaction is significant. Accordingly, we obtain
 \begin{equation}
 \delta \approx  \frac{ \nuDB}{ \nuSH^2}\left( \frac{\nuDB \taurInv}{ \Eceffeq/E_{\rm c}}+ \phibrI + \phibrII\ln\!\frac{\nuDB}{2\nuSH}\right).\label{eq:deltathen}
\end{equation}
Equation~\eqref{eq:EceffAnalyticalApp} is a not in a closed form since $\delta$
depends on \Eceff, but an accurate approximation is obtained after one iteration of~\eqref{eq:EceffAnalyticalApp} and~\eqref{eq:deltathen}. This is shown in a comparison with the full numerical solution to~\eqref{eq:Eceffeq} in figures~\ref{Eceff} and~\ref{fig:EceffApp}.

 \section*{References}
\bibliography{references} % references.bib

\end{document}